\documentclass[12pt]{article}
\def\baselinestretch{0.9}
\usepackage{anysize}
\usepackage{setspace}
\usepackage{times}
\usepackage{mathptm}
\usepackage{graphicx}
\marginsize{1in}{1in}{1in}{1in}

\usepackage{times,amssymb,mathptm,graphicx}

\newtheorem{lemma}{Lemma}
\newtheorem{proposition}{Proposition}
\newcommand{\vect}[1]{\mbox{$\underline{#1}$}}

\title{Fault Testing for Reversible Circuits\thanks{
\footnotesize{This work was supported by the DARPA QuIST program.
The views and conclusions contained herein are those of
the authors and should not be interpreted as necessarily
representing official policies of endorsements, either 
expressed or implied, of the Defense Advanced Research 
Projects Agency (DARPA) or the U.S. Government. A preliminary
version of this paper was presented at the VLSI Test Symposium 
, Napa, CA in April 2003.}}
}

\date{}
\author{Ketan N. Patel, John P. Hayes and Igor L. Markov\\
University of Michigan,
Ann Arbor 48109-2122 \\
{\tt \{knpatel,jhayes,imarkov\}@eecs.umich.edu}}

\begin{document}

\maketitle

\vspace{10mm}

\begin{abstract}
Applications of reversible circuits can be found in the fields of 
low-power computation, cryptography, communications, digital signal
processing, and the emerging field of quantum
computation. Furthermore, prototype circuits for low-power 
applications are already being fabricated in CMOS. Regardless of the
eventual technology adopted, testing is sure to be an important
component in any robust implementation.    

We consider the test set generation problem. Reversibility affects the
testing problem in 
fundamental ways, making it significantly simpler than for 
the irreversible case.  For example, we show that any test set
that detects all single stuck-at faults in a reversible circuit also
detects all multiple stuck-at faults. We present efficient test set
constructions for the standard stuck-at fault model as well as the
usually intractable cell-fault model.  We also give a practical test
set generation algorithm, based on an integer linear programming
formulation, that yields test sets approximately half
the size of those produced by conventional ATPG.  
\end{abstract}
\newpage 

%\begin{keywords}
%ATPG, reversible circuits, quantum computation, fault testing
%\end{keywords}
\section{Introduction}
A primary motivation for the study of reversible circuits is
the possibility of nearly energy-free computation. Landauer~\cite{landauer:iah:61} 
showed that traditional irreversible  circuits necessarily dissipate energy
due to the erasure of information. On the other hand, in principle,
reversible computation can be performed with arbitrarily small energy 
dissipation~\cite{bennett:lro:73,fredkin:cl:82}. Though the fraction of the
power consumption in current VLSI circuits attributable to information loss
is negligible, this is expected to change as increasing packing densities force
the power consumption per gate operation to
decrease~\cite{zhirnov:ltb:03}, making reversible 
computation an attractive alternative.

Many applications in the fields of cryptography, communications and
digital signal processing require computations that transform the data
without erasing any of the original information.  These applications are 
particularly well-suited to a reversible circuit implementation.
However, the applicability of reversible circuits is not limited to
inherently reversible applications.  Conventional irreversible
computation can be implemented reversibly using limited
overhead~\cite{li:rso:98,buhrman:tas:01}.  Furthermore, reversible
circuits are not just a theoretical area of study: De Vos et
al.~\cite{devos:arc:02} have built reversible CMOS circuits that are
powered entirely from the input pins without the
assistance of additional power supplies.

A major new motivation for the study of reversible circuits is provided
by the emerging field of quantum computation~\cite{nielsen:qca:00}. In a 
quantum circuit the 
operations are performed on quantum states or qubits rather than bits. 
Since quantum evolution is inherently reversible, the resulting quantum 
computation is as well. Classical reversible circuits form an important 
subclass of these quantum circuits. 
%Nielsen and Chuang~\cite{nielsen:qca:00} 
%provide an excellent background on quantum computation.

While logic synthesis, and hardware implementations for reversible
circuits have been studied in previous work, very little  
research has considered reversibility in the context of testing.
One exception is research at Montpellier, where 
reversibility was used to synthesize on-line test structures for
irreversible circuits~\cite{bertrand:slr:74,bertrand:slr2:74}. 
In contrast, our focus is on testing inherently reversible circuits,
particularly, generating efficient test sets for these circuits. 
Though this is a hard problem for conventional irreversible circuits,
it can be significantly simplified in  
our case. Agrawal~\cite{agrawal:ait:81} has shown that fault detection
probability is greatest when the information output of a circuit is 
maximized. This suggests that it may be easier to detect faults in reversible 
circuits, which are information lossless, than in 
irreversible ones. While this previous work focused on probabilistic 
testing, here we are concerned with complete deterministic testing.
We show that surprisingly few test vectors are necessary to
fully test a reversible circuit under the multiple stuck-at fault model, with 
the number growing at most 
logarithmically both in the number of inputs and the number of gates. 
This provides additional motivation for studying reversible circuits, namely 
they may be much easier to test than their irreversible counterparts.

In Section~\ref{sec:rev_circ} we give some basic background on reversible
circuits.  We then present some theoretical results on complete test sets
for reversible circuits in Section~\ref{sec:complete}, and results on
such sets for worst-case circuits in Section~\ref{sec:worst-case}.  In
Section~\ref{sec:ilp} we give a practical algorithm for generating
efficient complete test sets, present simulation results, and make
comparisons to conventional ATPG~(automatic test pattern generation).
We extend our results to the more general cell-fault model in
Section~\ref{sec:cell_fault}, and conclude in Section~\ref{sec:concl}.   

\section{Reversible Circuits}\label{sec:rev_circ}
\begin{figure}[t!]
\begin{center}
\resizebox{4.5in}{!}
{\includegraphics{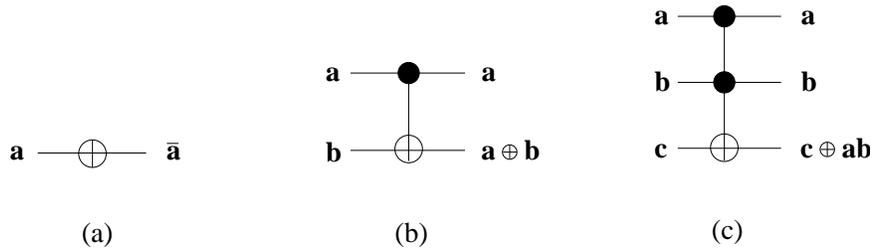}} \\
\caption{Examples of reversible logic gates: (a) NOT,  (b) C-NOT, and  (c) Toffoli.}\label{fig:cnt}
\end{center}
\end{figure}

A logic gate is \textit{reversible} if the mapping of inputs to outputs
is bijective, that is, every distinct input yields a distinct output, and
the numbers of input and output wires are equal. If it has $k$ inputs~(and outputs),
we call it a \textit{reversible $k\times k$ gate}. Three commonly used
gates, composing the $NCT$-gate 
library,  are shown in Figure~\ref{fig:cnt}. The NOT gate inverts the input, the 
C-NOT gate passes the first input through and inverts the second if the first 
is 1, and the Toffoli gate passes the first two inputs through and inverts the 
third if the first two are both 1.

A \textit{well-formed reversible circuit}
is constructed by starting with $n$ wires, forming the basic circuit, and iteratively
concatenating reversible gates to some subset of the output wires of the previous
circuit. The outputs of each reversible gate replace the wires at its input.
This iterative construction naturally gives us the 
notion of \textit{levels} in the circuit; the inputs to the circuit are at 
level 0, and the outputs of any gate are at one plus the highest level of any
of its inputs. For convenience in cases where a wire at the input of a gate
is at level $i$ and the outputs are at level $j>i+1$, we say the input is at
all levels between $i$ and $j-1$ inclusively. This gives us $n$ wires at each
level. Figure~\ref{fig:ex_circuit} shows an example of a reversible circuit
with the levels denoted by dotted lines.  The propagation of an input
vector through the circuit is shown to illustrate the circuit's operation.
The \textit{depth} $d$ of the circuit is the maximum level, which
can be no larger than the number of gates in the circuit.
We will often find it convenient to use an $n$-bit vector to refer to
the values of  
the wires at a given level in the circuit. A binary vector has \textit{weight} 
$k$ if it contains exactly $k$ 1's, and we denote the all-0's and all-1's 
vectors by $\vect{0}$ and $\vect{1}$, respectively.

The foregoing iterative construction also leads to the notion of a
\textit{sub-circuit}, the part of the original circuit between levels
$i$ and $j$, or more specifically, the circuit formed by the gates with
outputs at level greater than $i$ and less than $j+1$. We denote the function computed by the sub-circuit as $f_{i,j}$ 
and its
inverse as $f_{i,j}^{-1}$. If we omit the first subscript $i$ it should be
assumed to be $0$. The function of the entire circuit is then 
$f_d$.

We say a reversible circuit is \textit{$L$-constructible}, if it can be formed using 
the $L$-gate library.  Some important gate libraries used here are the 
$NCT$-gate library mentioned above, the C-NOT gate library consisting of only 
C-NOT gates, and the universal or $U$-gate library which consists of
all possible  
reversible $n\times n$ gates. These three gate libraries compute 
the set of even permutations, the set of linear reversible functions, and the
set of all permutations respectively~\cite{shende:sor:03}. In order to
compute any function that is not an even permutation, at least one
gate that spans all $n$ wires is required.  Unlike the $U$-library,
practical gate libraries are unlikely to contain such large gates. The
$NCT$-gate library has been well
studied~\cite{toffoli:rc:80,shende:sor:03} and 
computes essentially all functions that are practically
realizable. Consequently, we will focus on it for most of our work.

%\section{Fault Testing}
\section{Complete Test Sets}\label{sec:complete}
Given a reversible circuit $C$ and a fault set $F$,  we want 
to generate a set of test vectors that detect all faults in $F$.
We call such a test set \textit{complete}. A complete test set with the 
fewest possible vectors is \textit{minimal}.
\begin{figure}[t!]
\begin{center}
\resizebox{4.5in}{!}
{\includegraphics{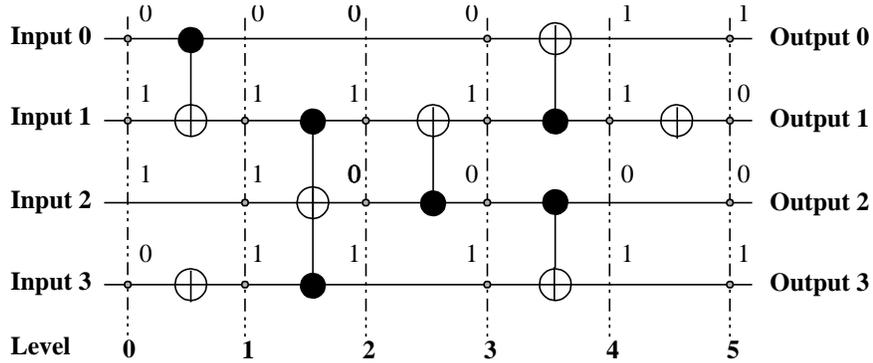}} \\
\caption{Reversible circuit example. The dotted lines represent the levels
in the circuit, and the small open dots represent possible stuck-at
fault sites.  The diagram also shows the propagation of input vector
$0110$ through the circuit.}\label{fig:ex_circuit} 
\end{center}
\end{figure}

Two important properties of reversibility simplify the 
test set generation problem. The first is \textit{controllability}: there 
is a test vector that generates any given desired state on the wires at 
any given level. The second is \textit{observability}: any single fault that 
changes an intermediate state in the circuit also changes the output. 
Neither property holds, in general, for irreversible circuits.  

To illustrate these two properties, consider the reversible circuit shown
in Figure~\ref{fig:ex_circuit}.  The controllability property enables
us to set the wires at any level in the circuit to any desired set of
values using a unique input vector, found by reversing the action of
the circuit.  For example, to find the input vector necessary to set
the wires at level 2 to the vector 0101, we first backtrack
through the three-input gate between levels 1 and 2.  This gives us
the vector 0111 at level 1. Backtracking once more gives the vector 0110 at
the input.  Reversibility guarantees that this backtracking is always
possible and always yields a unique vector at the input. The
observability property enables us to observe any 
intermediate change in the circuit at the output.  This is because
each vector at any level in the circuit corresponds to exactly one
output vector.  For example, only the vector 0101 at level 2 in the
circuit yields the output vector 1001;
any other vector at this level will yield a different output.

For most of this paper we adopt the standard stuck-at fault model 
used in testing conventional circuits, which includes all faults 
that fix the values of wires in the circuit to either $0$ or $1$.
For reversible circuits we show that any test set that detects all single 
stuck-at faults, also detects any number of simultaneous faults.
In Section~\ref{sec:cell_fault} we extend our results to the
more general cell-fault model, where the fault set consists of single
gate failures. 

\subsection{General Properties}
The following proposition provides a simple necessary and sufficient 
condition for a test set to be complete for the stuck-at fault model.
\begin{proposition}\label{prop:suff}
Under the single stuck-at fault model a test set is complete if and
only if each wire at every level can be set to both 0 and 1 by
the test set.  
\end{proposition}
\textbf{Proof}
Assume without loss of generality that a test set does not set a wire at 
level $i$ to $0$. A stuck-at $1$ fault at this point in the circuit is then
undetectable, since the outputs from the test set are unaffected.
On the other hand, if all wires at
every level can be set to both $0$ and $1$ by the test set, then a stuck-at fault must
affect at least one test vector, changing the value of the wire at that level from a $0$
to a $1$ or vice versa. By the observability property this change will affect the output.
$\Box$

\noindent 

To illustrate this proposition consider the fault site on the second wire at 
level 4 in the circuit in Figure~\ref{fig:ex_circuit}.  In order to detect a 
stuck-at 0 fault, the test set must be able to set this wire to 1, otherwise 
the fault would not have any effect on the test set, and would
therefore be undetected.  If a stuck-at 0 fault does occur, then a
test vector that sets the wire to 1 would generate an  
incorrect output, namely a 1 instead of a 0 on the second wire at
the output.  Similarly, to detect a  
stuck-at 1 fault on this wire, the test set must be able to set the wire to 0.

The next proposition shows that the single stuck-at and the multiple
stuck-at fault models are essentially identical for reversible 
circuits; specifically, a test set that is complete for one model is also
complete for the other. The intuition behind this property is that in
the case of multiple faults the final fault(s), i.e., those closest to
the outputs, can be detected by working backwards from the outputs.

\begin{proposition}\label{prop:suff_mult}
Any test set that is complete for the single stuck-at fault model
is also complete for the multiple stuck-at fault model. 
\end{proposition}
\textbf{Proof}
Suppose we have a counter-example. Then there must be a complete
test set $T$ for some reversible circuit under the single fault
model, which is not complete for multiple faults. So at least 
one multiple fault $M$ is undetectable by $T$. 
Since $M$ is undetectable, the response of the circuit to $T$ 
must be the same as those of the fault-free circuit. 
Now $M$ is composed of faults at various levels. Let $i$ be the deepest level
containing a sub-fault of $M$. Since no sub-faults occur at any level
greater than $i$, the reversible sub-circuit between level $i$ and the outputs
is identical to the corresponding sub-circuit in the fault-free circuit.
Therefore, since the response to $T$ and the reversible sub-circuit 
between level $i$ and the outputs are the same as for the fault-free circuit, 
the values of the wires at level $i$ must also be the same as for the 
fault-free circuit. 
Since $T$ is complete under the single
fault model, by Proposition~\ref{prop:suff} each wire at level $i$ must
take both the value 0 and 1. However this is a contradiction, since there
is at least one sub-fault at level $i$ that fixes the value of a wire.
$\Box$

This correspondence between the single and multiple stuck-at fault models 
allows us to restrict our attention to the conceptually simpler case
of single faults.    
If we have an $n$-wire circuit with $l$ gates of sizes 
$k_1,\ldots,k_l$, then a total of $2(n+\sum_{i=1}^{l}k_i)$
single stuck-at faults can occur: stuck-at 0 and stuck-at 1 faults 
for each gate input and circuit output. Reversibility then 
implies the following result, which will be useful later. 
\begin{lemma}\label{lemma:count}
Each test vector covers exactly half of the possible faults, and each
fault is covered by exactly half of the possible test vectors.
\end{lemma} 
\textbf{Proof}
Each test vector $t$ sets the bit at each fault site to either 0 or 1,
detecting either a stuck-at 1 or stuck-at 0 fault,
respectively. Therefore, $t$ detects precisely half of  the possible 
single stuck-at faults. For a given stuck-at fault there are $2^{n-1}$
possible bit vectors at that level that can detect the fault, namely
those that set the faulty bit to the opposite of the stuck-at
value. Since the circuit is reversible, each of these  
can be traced back to a distinct input vector. Therefore, half of the $2^n$ 
input vectors detect the fault.
$\Box$

We can obtain some properties of a minimal test set of a circuit by 
decomposing the circuit into sub-circuits. For example, the size of a minimal 
test set for a reversible circuit is greater than or equal to that of any of 
its sub-circuits. 
On the other hand, the size of a minimal test set for a circuit formed by 
concatenating reversible circuits $C_1$,$\ldots$, $C_k$ is no greater than the 
sum of the sizes of minimal test sets for the individual $C_i$'s. 
Finally if two reversible circuits $C_1$ and $C_2$, with minimal test
sets of sizes $|T_1|$ and $|T_2|$ respectively, act on a disjoint set
of input/output bits, then the size of the minimal test set of the
circuit formed by concatenating $C_1$ and $C_2$ is equal to
$\max\left\{|T_1|,|T_2|\right\}$.  These properties can be used to
bound the size of the minimal test set, and in some cases, to simplify
the problem of finding a minimal test set. 

\subsection{Test Set Construction}
The following proposition gives a number of complete test set constructions,
implicitly providing upper bounds on the size of a minimal test set.
\begin{proposition}\label{prop:constr}
A complete test set for an $n$-wire reversible circuit with depth $d$ and a total of $l$ gates with 
sizes $k_1,\ldots,k_l$ is given by:
\begin{itemize}
\item[a.] any $2^{n-1}+1$ distinct test vectors
\item[b.] the following $d+2$ test vectors
\begin{equation}
\left\{\vect{0},\ \vect{1},\ f_{1}^{-1}(\overline{f_1(\vect{0})}),\ldots,\ 
f_d^{-1}(\overline{f_d(\vect{0})})\right\}
\end{equation}
%where $\vect{0}$ and $\vect{1}$ are the all zeros and all ones vectors respectively. 
\item[c.] some set of $\left\lfloor\log_2\left(n+\sum_{i=1}^{l}k_i \right)\right\rfloor+2$ test vectors.
\end{itemize}
\end{proposition}
\textbf{Proof}\\
\noindent(a) The value of a wire at a given level is set to $0$ (or $1$) by exactly $2^{n-1}$ input
vectors. Therefore, if the test set contains $2^{n-1}+1$ vectors, then at least one 
will set it to $1$ (or $0$). Since this is true for all fault sites, by 
Proposition~\ref{prop:suff} the test set is complete.

\noindent(b) The vector
$f_{i}^{-1}(\overline{f_i(\vect{0})})$
%$\vect{1}, \  f_{1}^{-1}(\overline{f_1(\vect{0})}),\ \ldots,
%\  f_d^{-1}(\overline{f_d(\vect{0})})$ 
sets the wires at level $i$
to the bitwise inverse of the values set by the $\vect{0}$
vector. Therefore each wire at every level can be set to both $0$ and
$1$ by the test set. By Proposition~\ref{prop:suff} the test set is complete.

\noindent(c) To prove this part we first prove that given a reversible circuit and 
an incomplete set of test vectors, there is a test vector that can be added that 
covers at least half of the remaining faults.

Let $m$ be the number of test vectors given, $F_C$ be the faults covered by 
this set, and $C$ the size of $F_C$. If none of the remaining $2^n-m$ input vectors 
cover at least half of the remaining faults, then they must each cover more than half of 
the faults in $F_C$. By Lemma~\ref{lemma:count} every test vector covers exactly 
$n+\sum_{i=1}^{l}k_i$ faults and every fault is covered by exactly $2^{n-1}$ test vectors. 
Therefore, the number of times faults in $F_C$ are covered by all input 
vectors cumulatively is $2^{n-1}\cdot C$, implying the following 
inequalities:
\begin{eqnarray}
%\left(2^n-m\right)\left(\left\lceil\frac{C}{2}\right\rceil\right)&\le& 2^{n-1}\cdot C -m\left(n+\sum_{i=1}^{l}k_i\right) \\
\left(2^n-m\right)\left(\frac{C}{2}\right)&<& 2^{n-1}\cdot C-m\left(n+\sum_{i=1}^{l}k_i\right) \\
2\left(n+\sum_{i=1}^{l}k_i\right) &<& C
\end{eqnarray}
The second inequality is false since the number of faults covered cannot be larger
than the total number of faults that can occur.
Therefore we have a contradiction, and there must be a test vector
that can be added to cover at least half of the remaining faults.

Recursively applying this observation we can eliminate all uncovered
faults in no more than
%reduce the number of uncovered 
%faults to zero in no more than 
\begin{equation}
\left\lfloor \log_2 \left(n+\sum_{i=1}^{l}k_i\right)\right\rfloor +
2
\end{equation}
steps (test vectors).
%$2n(d+1)\rightarrow n(d+1)\rightarrow \left\lfloor n(d+1)/2\right\rfloor \rightarrow \cdots \rightarrow 0$, 
$\Box$

Proposition~\ref{prop:constr} limits the size of the minimal test set based 
on the size of the reversible circuit both in terms of its depth 
and the number of input/output bits. For the circuit in 
Figure~\ref{fig:ex_circuit}, parts a-c of the proposition give upper bounds 
of 9, 7, and 6 test vectors, respectively. The final part of 
the proposition implies that a reversible circuit can be tested by a very
small set of tests.
As an example, a reversible circuit on 64 wires with a million
$3\times 3$ gates can be tested using no more than 23 input vectors.
However, while the first two parts of Proposition~\ref{prop:constr}
give practical constructions, the last one does not; consequently, it
may not be easy to find such a test set. 

\section{(L,n)-Complete Test Sets}\label{sec:worst-case}
We say a test set is \textit{$(L,n)$-complete} for gate library $L$ 
acting on $n$ wires, if it is complete for all circuits formed by the library. 
The following proposition shows that a circuit requiring such a 
test set exists for any gate library.
\begin{proposition}
Any reversible gate library $L$ acting on $n$ wires has an $(L,n)$-complete 
set of test vectors that is minimal for some circuit in the set.
\end{proposition}
\textbf{Proof}
Let $C_1,\ldots,C_N$ be a set of circuits that computes the set of
all functions computable using $L$, and 
$C=C_1C_1^{-1}\cdots C_N C_N^{-1}$. Then any test set that is 
complete for $C$ must be complete for any circuit formed by $L$. Therefore,
a minimal test set for $C$ is $(L,n)$-complete. 
$\Box$

The following proposition characterizes $(L,n)$-complete test sets for three 
classes of reversible circuits: $C$-constructible, $U$-constructible, 
and $NCT$-constructible. 
\begin{proposition}$\ $
\begin{itemize}
\item[a.] A $(C,n)$-complete test set must have at least $n+1$ vectors. One
such set comprises the all-0's vector and the $n$ weight-1 vectors.
\item[b.] A $(U,n)$-complete test set must have at least $2^{n-1}+1$ vectors,
and any $2^{n-1}+1$ test vectors will give such a set.
\item[c.] An $(NCT,n)$-complete test set must have at least $2^{n-1}+1$ vectors,
and any $2^{n-1}+1$ test vectors will give such a set.
\end{itemize}
\end{proposition}
\textbf{Proof}\\
\noindent(a) Any input to the circuit can be written as a linear combination
of the $n$ weight-1 vectors. Furthermore, since the gate library is 
linear~(under the operation of bitwise XOR), 
the corresponding values of the wires at the $i$th level can be written as the
same linear combination of the values for these weight-1 vectors.
If any input vector sets the value of a wire at the $i$th level to 1, then 
so must at least one weight-1 vector. Since there are inputs that do,
the weight-1 vectors are sufficient for setting all wires to 1. Furthermore,
since the circuit is linear, the all-0's vector sets all wires at all levels 
to 0. Therefore, this is a $(C,n)$-complete test set. In general any $n$ 
linearly independent vectors along with the all-0's vector forms a 
$(C,n)$-complete test set.

On the other hand, if the test set consists of only $n$ input vectors, we have 
two possibilities: either the set spans the $n$-dimensional space or it does
not. If the latter case, a linear reversible circuit can be constructed 
that maps the test set into the $(n-1)$-dimensional subspace $0X\cdots X$, 
implying that the test set is not complete. If the test set spans the
entire $n$-dimensional space, a linear reversible circuit can be constructed 
that maps them to the linearly independent vectors:
\[\renewcommand{\arraycolsep}{.04in}
\begin{array}{ccccccc}
v_1 \rightarrow & 1 & 0 & 0 & 0 & \cdots & 0 \\
v_2 \rightarrow & 1 & 1 & 0 & 0 & \cdots & 0 \\
v_3 \rightarrow & 1 & 1 & 1 & 0 & \cdots & 0 \\
\vdots  & & \vdots &  & \vdots & & \vdots \\
v_n \rightarrow & 1 & 1 & 1 & 1 & \cdots & 1
\end{array}
\]
Since the first wire cannot be set to 0, the test set is not complete for this
circuit.

\noindent(b) Suppose we have a $(U,n)$-complete test set with $2^{n-1}$ test vectors.
Because the gate library computes all permutations, we can generate a circuit 
mapping all $2^{n-1}$ test vectors to output vectors of the 
form $0\mbox{XX}\cdots \mbox{X}$. This test set does not set the first output 
bit to $1$, and thus is not complete for this $U$-gate circuit. This implies 
it is not $(U,n)$-complete.
By Proposition~\ref{prop:constr}a, any $2^{n-1}+1$ test vectors will give
$(U,n)$-completeness.

\noindent(c) Any permutation can be composed from a series of transpositions.
The $NCT$ gate library can construct circuits computing any even permutation 
of the input values~\cite{shende:sor:03}, that is, a permutation that can be composed from an even 
number of transpositions. Following the proof for part b, a permutation can
map any $2^{n-1}$ test vectors to output vectors of the form $0XX\cdots X$. 
If this permutation is even we have shown that this is an incomplete test set,
otherwise we can add a transposition that exchanges the outputs $00\cdots 0$ 
and $00\cdots 1$. This new permutation is even and still maps the test vectors
to the set of outputs $0XX\cdots X$, and therefore, the test set is not complete
for this $NCT$-circuit. By Proposition~\ref{prop:constr}a, any $2^{n-1}+1$ test 
vectors will give $(NCT,n)$-completeness.
$\Box$

Note that any two gate libraries that can compute the same set of 
functions are equivalent with respect to $(L,n)$-completeness.  This is because
the function of every gate of one library can be computed by the other.  
Therefore, if a test set is not $(L,n)$-complete for one library, it cannot
be for the other either, implying that the two libraries share 
the same $(L,n)$-complete test sets.  This means that the above result for 
the $U$-gate library is applicable to any library that can compute all 
$n$-bit reversible functions.

$C$-constructible circuits are analogous to XOR-trees, since C-NOT
gates are simply XOR gates with an additional output that is equal to
one of the inputs. Consequently, part (a) of the above proposition can
be considered the reversible 
analog of the well-known result that any XOR-tree can be tested for
single stuck-at faults using no more than four tests~\cite{hayes:oro:71}.
We consider linear reversible circuits separately here, primarily 
because they can be tested with a very simple set of tests just as XOR-trees
in conventional irreversible circuit testing~\cite{debany:afg:91}.

\section{ILP formulation}\label{sec:ilp}
While Proposition~\ref{prop:constr}c guarantees that an efficient test set
exists for any reversible circuit, it gives no practical construction.
In this section, we formulate the problem of constructing a minimal test 
set as an integer linear program~(ILP) with binary variables. We then use 
this to find a practical heuristic for generating efficient test sets. 

% Alternatively, we could formulate the problem as an instance of 
% \textit{satisfiability}~(SAT)~\cite{garey:79:cai}. This is 
% distinct from the SAT formulation
% of the ATPG~(automatic test pattern generation)
% problem~\cite{stephan:ctg:96}, 
% where the object is to find a test pattern that detects a given 
% fault or prove the fault undetectable. In our case the ATPG problem is relatively easy because of 
% reversibility; the difficult part is finding a minimal test set.
\subsection{ILP Model}\label{subsec:ilp}
We can formulate the minimal test set problem as an ILP with
binary decision variables $t_i$ associated with each input vector $T_i$; $t_i$ takes a value 
of one if the corresponding input vector is in the test set, and zero otherwise. A 
fault is detected if a decision variable with value one corresponds to a vector
that detects the fault.  The values of the wires at the $j$-th level for input $T_i$ are 
$f_j\left(T_i\right)$, so to detect all stuck-at 0 faults at level $j$ the following 
inequalities must be satisfied
\[
\sum_{i=0}^{2^n-1} f_j\left(T_i\right)\cdot t_i \ge \underline{1}
\] 
These inequalities guarantee that each wire at the $j$-th level is set
to 1 by some test vector, ensuring that all stuck-at 0 faults are
detected. A similar set of inequalities ensures that all stuck-at 1
faults are also detected.  
In total, $2n(d+1)$ linear inequality constraints are used to guarantee
completeness. We determine a minimal test set by minimizing the sum
of the $t_i$'s, that is, by minimizing the size of the complete test set,
subject to these constraints.  The general ILP formulation of  
the minimal test set problem for a reversible circuit on $n$ wires
and with depth $d$ is:
\begin{eqnarray*}
%\begin{array}{lll}
&&\mbox{\textbf{Minimize} } t_0+t_1+\cdots+t_{2^n-1} \\
&&\mbox{\textbf{subject to the constraints}} \\
&&\ \ \ \ \ \sum_{i=0}^{2^n-1} f_j\left(T_i\right)\cdot t_i \ge \underline{1} \ \ \\
&&\ \ \ \ \ \sum_{i=0}^{2^n-1} \overline{f_j\left(T_i\right)}\cdot t_i \ge \underline{1},\ \ \ \ \ 
\mbox{for all } 0\le j \le d \\
&&\mbox{where }t_i\in \left\{0,1\right\},\ \mbox{and } \\
&& T_i \mbox{ is the } n\mbox{-bit binary expansion of integer }i\ \ \ \ \ \ \ \ \ \ \ \ \ \ 
%  \left[\begin{array}{cccc}
%    \overline{T_0} & \overline{T_1} &  \cdots &  \overline{T_{2^n-1}}\\
%    T_0 &  T_1 &  \cdots &  T_{2^n-1}\\
%    \overline{f_1(T_0)} &  \overline{f_1(T_1)} &  \cdots &  \overline{f_1(T_{2^n-1})} \\
%    f_1(T_0) &  f_1(T_1) &  \cdots &  f_1(T_{2^n-1}) \\
%    \vdots & \vdots &  \vdots &  \vdots \\
%    \overline{f_d(T_0)} &  \overline{f_d(T_1)} &  \cdots &  \overline{f_d(T_{2^n-1})} \\
%    f_d(T_0) &  f_d(T_1) & \cdots &  f_d(T_{2^n-1}) \\
%    \end{array}\right]\cdot\left[\begin{array}{c}
%    t_0\\ t_1\\ \vdots \\ t_{2^n-1}\end{array}\right] \ge \vect{1}\\ \\
%\end{array}
\end{eqnarray*}

%In this ILP $t_0,\ldots,\ t_{2^n-1}$ are indicator variables each representing
%the $2^n$ possible test vectors, and each of the 
%$2n(d+1)$ linear constraints ensures that the test set can set a wire in the circuit to 
%either 0 or 1. 
\noindent Each feasible solution gives  a complete test set composed of those 
vectors $i$ for which $t_i=1$. For relatively small circuits 
this ILP can be solved efficiently with an off-the-shelf optimization tool 
such as CPLEX~\cite{cplex}.  
\begin{figure}[t!]
\begin{center}
\resizebox{1.25in}{!}
{\includegraphics{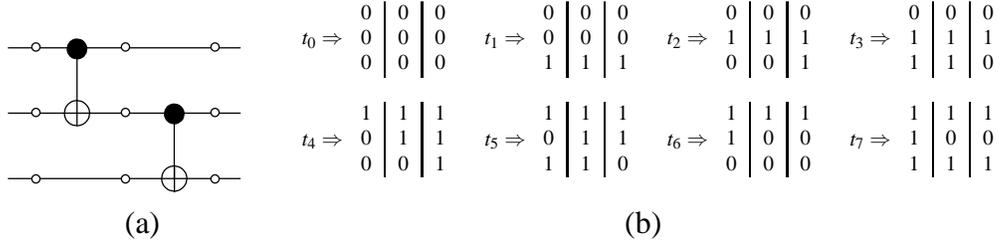}} 
$\scriptsize\ \ \ \ \ \ \ \begin{array}[b]{cccc}
t_0 \Rightarrow \begin{array}{c|c|c} 0 & 0 & 0\\ 0 & 0 & 0\\ 0 &
  0 & 0\end{array} & 
t_1 \Rightarrow \begin{array}{c|c|c} 0 & 0 & 0\\ 0 & 0 & 0\\ 1 &
  1 & 1\end{array} & 
t_2 \Rightarrow \begin{array}{c|c|c} 0 & 0 & 0\\ 1 & 1 & 1\\ 0 &
  0 & 1\end{array} & 
t_3 \Rightarrow \begin{array}{c|c|c} 0 & 0 & 0\\ 1 & 1 & 1\\ 1 &
  1 & 0\end{array} \\ \\
t_4 \Rightarrow \begin{array}{c|c|c}
  1 & 1 & 1\\ 0 & 1 & 1\\ 0 & 0 & 1\end{array} &
t_5 \Rightarrow \begin{array}{c|c|c} 1 & 1 & 1\\ 0 & 1 & 1\\ 1 &
  1 & 0\end{array} & 
t_6 \Rightarrow \begin{array}{c|c|c} 1 & 1 & 1\\ 1 & 0 & 0\\ 0 &
  0 & 0\end{array} & 
t_7 \Rightarrow \begin{array}{c|c|c}
  1 & 1 & 1\\ 1 & 0 & 0\\ 1 & 1 & 1\end{array} \\ \\
\end{array}$\\ 
\hspace{-1.25in}(a) \hspace{2.35in} (b)
\caption{(a) Reversible circuit example. Possible stuck-at fault sites
  are represented by small open dots. (b) Propagation of each of
  the possible input vectors through the circuit.  A complete test set
  must set each of the nine fault sites to both 0 and 1. An example of
  a complete test set for this circuit is $\{t_0,\ t_2,\ t_7\}$}\label{fig:ex_circuit2} 
\end{center}
\end{figure}

Consider the circuit shown in Figure~\ref{fig:ex_circuit2}.  
The corresponding ILP formulation is: 
\begin{eqnarray*}
%\begin{array}{lll}
&&\mbox{\textbf{Minimize} } t_0+t_1+t_2+t_3+t_4+t_5+t_6+t_7 \\
&&\mbox{\textbf{subject to the constraints}} \\
&&\begin{array}{ccc}
\begin{array}{c}
\sum_{i=0}^{7} f_0\left(T_i\right)\cdot t_i \ge \underline{1} \\ \\
\sum_{i=0}^{7} f_1\left(T_i\right)\cdot t_i \ge \underline{1} \\ \\
\sum_{i=0}^{7} f_2\left(T_i\right)\cdot t_i \ge \underline{1} \\
\end{array}
&\Longleftrightarrow &
\left[\renewcommand{\arraystretch}{.6}{\begin{array}{ccccccccc}
0&0&0&0&1&1&1&1\\
0&0&1&1&0&0&1&1\\ \vspace{.07in}
0&1&0&1&0&1&0&1\\
0&0&0&0&1&1&1&1\\
0&0&1&1&1&1&0&0\\ \vspace{.07in}
0&1&0&1&0&1&0&1\\
0&0&0&0&1&1&1&1\\
0&0&1&1&1&1&0&0\\
0&1&1&0&1&0&0&1
\end{array}}\right]
\cdot 
\left[\renewcommand{\arraystretch}{.6}{\begin{array}{c}
t_0\\
t_1\\ \vspace{.05in}
t_2\\
t_3\\
t_4\\ \vspace{.05in}
t_5\\
t_6\\
t_7\\
\end{array}}\right]
 \ge 
\left[\renewcommand{\arraystretch}{.6}{\begin{array}{c}
1\\1\\ \vspace{.07in}
1\\1\\1\\ \vspace{.07in}
1\\1\\1\\1
\end{array}}\right]\\ \\
\begin{array}{c}
\sum_{i=0}^{7} \overline{f_0\left(T_i\right)}\cdot t_i \ge
\underline{1}\\ \\
\sum_{i=0}^{7} \overline{f_1\left(T_i\right)}\cdot t_i \ge
\underline{1}\\ \\ 
\sum_{i=0}^{7} \overline{f_2\left(T_i\right)}\cdot t_i \ge
\underline{1}\\ 
\end{array}
&\Longleftrightarrow &
\left[\renewcommand{\arraystretch}{.6}{\begin{array}{ccccccccc}
1&1&1&1&0&0&0&0\\
1&1&0&0&1&1&0&0\\ \vspace{.07in}
1&0&1&0&1&0&1&0\\ 
1&1&1&1&0&0&0&0\\
1&1&0&0&0&0&1&1\\ \vspace{.07in}
1&0&1&0&1&0&1&0\\ 
1&1&1&1&0&0&0&0\\
1&1&0&0&0&0&1&1\\
1&0&0&1&0&1&1&0
\end{array}}\right]\cdot 
\left[\renewcommand{\arraystretch}{.6}{\begin{array}{c}
t_0\\
t_1\\ 
t_2\\
t_3\\
t_4\\
t_5\\
t_6\\
t_7\\
\end{array}}\right]
 \ge 
\left[\renewcommand{\arraystretch}{.6}{\begin{array}{c}
1\\1\\ \vspace{.07in}
1\\1\\1\\ \vspace{.07in}
1\\1\\1\\1
\end{array}}\right]\end{array}\\
% &(0)&\ \ \ \ t_0+t_1+t_2+t_3 \ge 0
% \ \ \ \ \ \ t_4+t_5+t_6+t_7 \ge 0\\
% &(1)&\ \ \ \ t_0+t_1+t_4+t_5 \ge 0
% \ \ \ \ \ \ t_2+t_3+t_6+t_7 \ge 0\\
% &(2)&\ \ \ \ t_0+t_2+t_4+t_6 \ge 0
% \ \ \ \ \ \ t_1+t_3+t_5+t_7 \ge 0\\
% &(3)&\ \ \ \ t_0+t_1+t_2+t3 \ge 0
% \ \ \ \ \ \ t_4+t_5+t_6+t_7 \ge 0\\
% &(4)&\ \ \ \ t_0+t_1+t_6+t7 \ge 0
% \ \ \ \ \ \ t_2+t_3+t_4+t_5 \ge 0\\
% &(5)&\ \ \ \ t_0+t_2+t_4+t_6 \ge 0
% \ \ \ \ \ \ t_1+t_3+t_5+t_7 \ge 0\\
% &(6)&\ \ \ \ t_0+t_1+t_2+t3 \ge 0
% \ \ \ \ \ \ t_4+t_5+t_6+t_7 \ge 0\\
% &(7)&\ \ \ \ t_0+t_1+t_6+t7 \ge 0
% \ \ \ \ \ \ t_2+t_3+t_4+t_5 \ge 0\\
% &(8)&\ \ \ \ t_0+t_3+t_5+t6 \ge 0
% \ \ \ \ \ \ t_1+t_2+t_4+t_7 \ge 0\\
&&\mbox{where }t_i\in \left\{0,1\right\}
\end{eqnarray*}
The first set of inequalities guarantees that each of the fault sites
can be set to 1, and the second set guarantees that they can be set to
0.  Solving the ILP, we find that three test vectors are required to
detect all stuck-at faults in the circuit.  One such solution is
$t_0=t_7=t_2=1$. 

Using the ILP formulation and CPLEX 7.0, we obtained minimal test sets for all 
optimal 3-wire $NCT$-circuits. CPLEX solved the ILP for each
circuit in a fraction of a second on a Sun SPARC. Table~\ref{tab:ilp}
gives a distribution of minimum test set size with respect to the
number of gates in the circuit. The optimal $NCT$
implementation of a given function is not unique, and therefore the
distribution in Table~\ref{tab:ilp} may be dependent on the particular
optimal set chosen. 
\begin{table} \small
\begin{center} \raisebox{-6.4ex}{\rotatebox{90}{\bf Test Size}}
\begin{tabular}{|c||c|c|c|c|c|c|c|c|c|c|}
\multicolumn{9}{c}{\bf Circuit Length (gates)} \\
\hline 
& \bf 0 & \bf 1 & \bf 2 &  \bf 3 &  \bf  4 &  \bf  5 &  \bf
6 & \bf 7  & \bf 8\\\hline \hline 
\bf 2 &  1 & 6 & 24 &   67 &  134 &  155 &   105 & 21  &  -\\\hline
\bf 3 &  - &  6 &  78 &  558 &  2641 &  8727 & 16854 &
10185  & 577 \\\hline 
\bf 4 &  - &  - & -  &  -  &     5 &   39 &    90 & 47  & -  \\\hline
\end{tabular}\caption{Minimal test set size distribution for optimal 3-wire 
$NCT$-circuits as a function of circuit length.}\label{tab:ilp}
\end{center}
\end{table}

As expected, the size of the minimal test set generally increases 
with the length of the circuit. On the other hand, there are long
circuits that have smaller minimal test sets than much shorter
circuits. The largest minimal test set has 4 vectors; however,
suboptimal circuits requiring 5 test vectors can be constructed.

\subsection{Circuit Decomposition Approach}\label{subsec:circ_decomp}
Solving the ILP exactly is feasible for small circuits; however, since the number of 
variables increases exponentially with the number of input/output bits, it
is impractical for large circuits.
An alternative approach is to decompose the original circuit into smaller sub-circuits 
acting on fewer input/output bits, and use the ILP formulation iteratively for these 
sub-circuits combining the test vectors dynamically;
a similar approach has been used for irreversible circuits~\cite{goel:tgd:79}.
While the resulting test set is not 
guaranteed to be minimal, it is generally small enough to enable efficient 
testing. Furthermore, it may be possible to use lower bounds to ensure the 
test set is not much larger than a minimal one. For example, the size of the
minimal test set of a sub-circuit can be used to bound that of the larger 
circuit.

The algorithm shown in Figure~\ref{fig:alg} uses this decomposition
approach. The circuit is first decomposed into a series of circuits
acting on a smaller number of wires. One way to do this is to start at
the input of the circuit, and add gates to the first sub-circuit $C_0$
until no more can be added without having $C_0$ act on more than $m$
wires. This is continued  with $C_1$, and so on until the entire circuit
has been decomposed. The remaining steps in the algorithm are best 
illustrated by an example.
\begin{figure}[t!]
\begin{center}
{\ttfamily \small
\begin{tabular}{|ll|}
\hline &\\
1) & Partition circuit into disjoint sub-circuits\Large{ } \\
& $C_0,\ldots,C_l$ each acting on $\le m$ wires\\
2) & Initialize test\_set $= \{\}$ and $i=0$\\
3) & Generate ILP for $C_i$ as in Section~\ref{subsec:ilp}\\
4) & Add constraints for each vector in test\_set \\
5) & Solve ILP\\
6) & Incorporate new test vectors into test\_set,\\
& setting any unused wires of new vectors to \\
& don't cares\\
7) & Apply $C_i$ to test\_set, setting don't cares at\\
&  inputs of $C_i$ to 0\\
8) & If $i<l$, $i=i+1$ and go to Step 3\\
9) & Set remaining don't cares in test\_set to 0\\
10) & Apply $C^{-1}$ to test\_set to get complete test set\\ & \\
\hline 
\end{tabular}
}
\caption{Algorithm for complete test set generation based on circuit decomposition.}\label{fig:alg}
\end{center}
\end{figure}

\begin{figure}[t!]
\begin{center}
\resizebox{3.75in}{!}
{\includegraphics{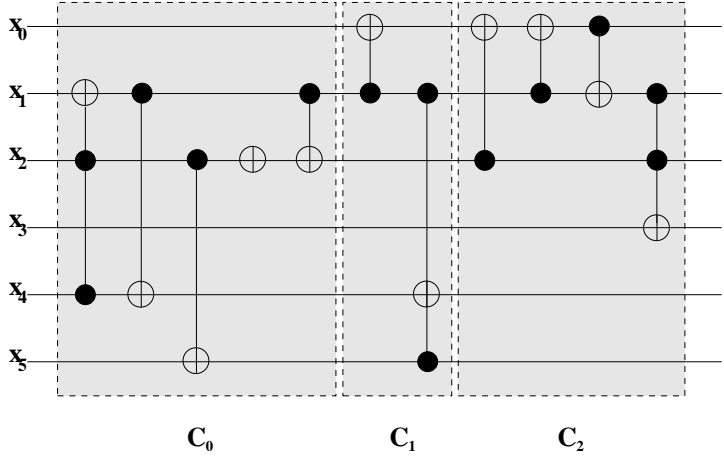}} \\
\caption{Circuit decomposition example.}\label{fig:decomp_ex}
\end{center}
\end{figure}

%\noindent\textbf{Example}
Consider the decomposition of the reversible circuit in 
Figure~\ref{fig:decomp_ex}. Though the entire circuit acts on six wires, 
each sub-circuit acts on no more than four. Using the ILP formulation on 
$C_0$ gives test vectors:
\[ \renewcommand{\arraycolsep}{.04in}\begin{array}{ccccccc}
& x_0 & x_1 & x_2 & x_3 & x_4 & x_5\\ \cline{2-7}
v_0= & \mbox{X} & 0 & 1 & \mbox{X} & 1 & 1\\  
v_1= &\mbox{X} & 1 & 0 & \mbox{X} & 0 & 0\\  
v_2= &\mbox{X} & 1 & 1 & \mbox{X} & 1 & 0
\end{array}
\stackrel{C_0}{\stackrel{}{\Longrightarrow}}
\begin{array}{cccccc}
x_0 & x_1 & x_2 & x_3 & x_4 & x_5\\ \hline 
\mbox{X} & 1 & 1 & \mbox{X} & 0 & 0\\  
\mbox{X} & 1 & 0 & \mbox{X} & 1 & 0\\  
\mbox{X} & 0 & 0 & \mbox{X} & 1 & 1
\end{array}
\]
where the X's represent don't cares and the left and right halves represent
the test vectors at the input and output of $C_0$, respectively. Sub-circuit 
$C_1$ acts on wires $x_0$, $x_1$, $x_4$ and $x_5$. We generate the ILP for 
$C_1$, and add the following constraints:
\[ \renewcommand{\arraycolsep}{.04in}\begin{array}{cccc}
x_0 & x_1 & x_4 & x_5  \\\hline
\mbox{X} & 1 & 0 & 0 \\
\mbox{X} & 1 & 1 & 0 \\
\mbox{X} & 0 & 1 & 1 \\
\end{array}
\in T \Rightarrow 
\begin{array}{lllll}
\multicolumn{5}{c}{\mbox{Constraints}} \\\hline
t_4 & + & t_{12} & \ge & 1 \\
t_6 & + & t_{14} & \ge & 1 \\
t_3 & + & t_{11} & \ge & 1 \\
\end{array}
\]
Solving this ILP gives the solution $t_6=t_{11}=t_{12}=1$. Incorporating
these values into the previous test vectors we have:
\[ \renewcommand{\arraycolsep}{.04in}
\begin{array}{cccccc}
x_0 & x_1 & x_2 & x_3 & x_4 & x_5\\\hline
1 & 1 & 1 & \mbox{X} & 0 & 0\\  
0 & 1 & 0 & \mbox{X} & 1 & 0\\  
1 & 0 & 0 & \mbox{X} & 1 & 1
\end{array}
\stackrel{C_1}{\stackrel{}{\Longrightarrow}}
\begin{array}{cccccc}
x_0 & x_1 & x_2 & x_3 & x_4 & x_5\\\cline{1-6}
0 & 1 & 1 & \mbox{X} & 0 & 0\\  
1 & 1 & 0 & \mbox{X} & 1 & 0\\  
1 & 0 & 0 & \mbox{X} & 1 & 1
\end{array} 
\]
Sub-circuit $C_2$ acts on wires $x_0$, $x_1$, $x_2$, and $x_3$. We generate
the ILP for this sub-circuit, and incorporate the current test set using
the following constraints:
\[\renewcommand{\arraycolsep}{.04in}\begin{array}{cccc}
x_0 & x_1 & x_4 & x_5  \\\hline
0 & 1 & 1 & \mbox{X} \\
1 & 1 & 0 & \mbox{X} \\
1 & 0 & 0 & \mbox{X} \\
\end{array}
\in T \Rightarrow 
\begin{array}{lllll}
\multicolumn{5}{c}{\mbox{Constraints}} \\\hline
t_{6} & + & t_{7} & \ge & 1 \\
t_{12} & + & t_{13} & \ge & 1 \\
t_8 & + & t_{9} & \ge & 1 \\
\end{array}
\]
Solving this ILP gives solutions $t_5$, $t_7$, $t_8$, and $t_{12}$. The last
three can be incorporated into the previous test set, however the first test
vector must be added:
\[\renewcommand{\arraycolsep}{.04in}
\begin{array}{cccccc}
x_0 & x_1 & x_2 & x_3 & x_4 & x_5\\\cline{1-6}
0 & 1 & 1 & 1 & 0 & 0\\  
1 & 1 & 0 & 0 & 1 & 0\\  
1 & 0 & 0 & 0 & 1 & 1\\
0 & 1 & 0 & 1 & \mbox{X} & \mbox{X}
\end{array} 
\stackrel{C_2}{\stackrel{}{\Longrightarrow}}
\begin{array}{cccccc}
x_0 & x_1 & x_2 & x_3 & x_4 & x_5\\\cline{1-6}
0 & 1 & 1 & 0 & 0 & 0\\  
0 & 1 & 0 & 0 & 1 & 0\\  
1 & 1 & 0 & 0 & 1 & 1\\
1 & 0 & 0 & 1 & \mbox{X} & \mbox{X}
\end{array} 
\]
Filling the don't cares with 0's and applying $C^{-1}$ to the test set yields
a complete test set for $C$. While the resulting test set is not guaranteed to be 
minimal, in this case it is, as can be shown by applying the ILP method on the 
entire circuit.

\subsection{Test Set Compaction} \label{subsec:test_comp}
The circuit decomposition method in the previous section generally produces
redundant test sets. One way to reduce this redundancy is to compact the
test set, that is, find the smallest complete subset. 
This approach has been used previously in ATPG algorithms for conventional  
circuits~\cite{hochbaum:aot:96,chang:tsc:95,flores:oas:99}.
%irreversible circuits~\cite{hochbaum:aot:96,chang:tsc:95,flores:oas:99}.
The ILP formulation in Section~\ref{subsec:ilp} can be used to perform the
test set compaction. We simply eliminate all test vectors that are not 
in the original complete test set, along with
the corresponding columns in the constraint matrix. Generally, this ILP can 
be solved more efficiently than the ILP for the minimal test set, since it has 
fewer variables.  Consider the example in the previous section. Since
the circuit decomposition method yields a complete test set with four test
vectors the ILP formulation for the test set compaction problem only
requires four variable, significantly less than the 64 required in the ILP
formulation for the minimal test set problem.

\subsection{Simulation Results}
\begin{figure}[t!]
\begin{center}
$\begin{array}{cc}
\resizebox{3.1in}{!}
{\includegraphics{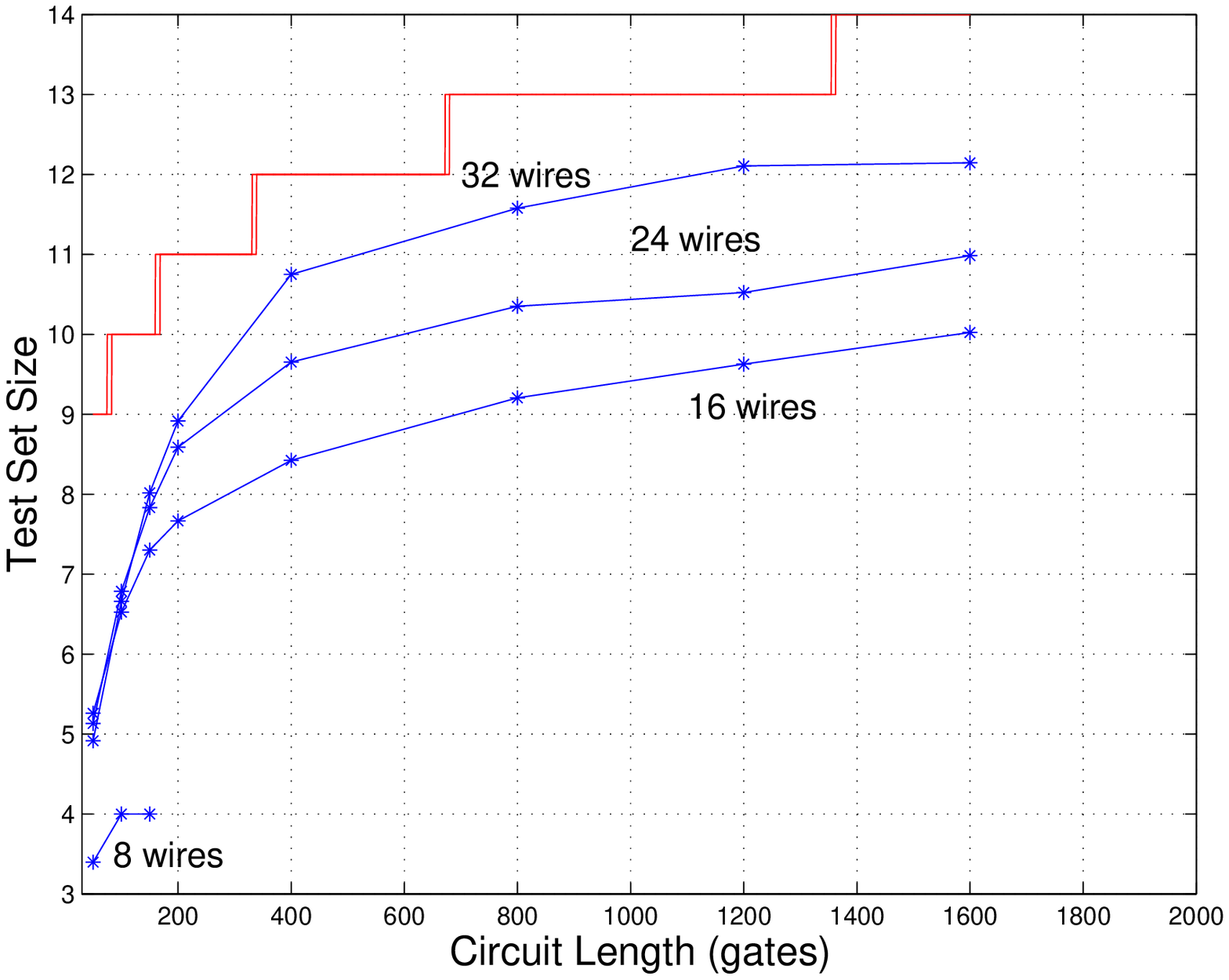}} &
\resizebox{3.1in}{!}
{\includegraphics{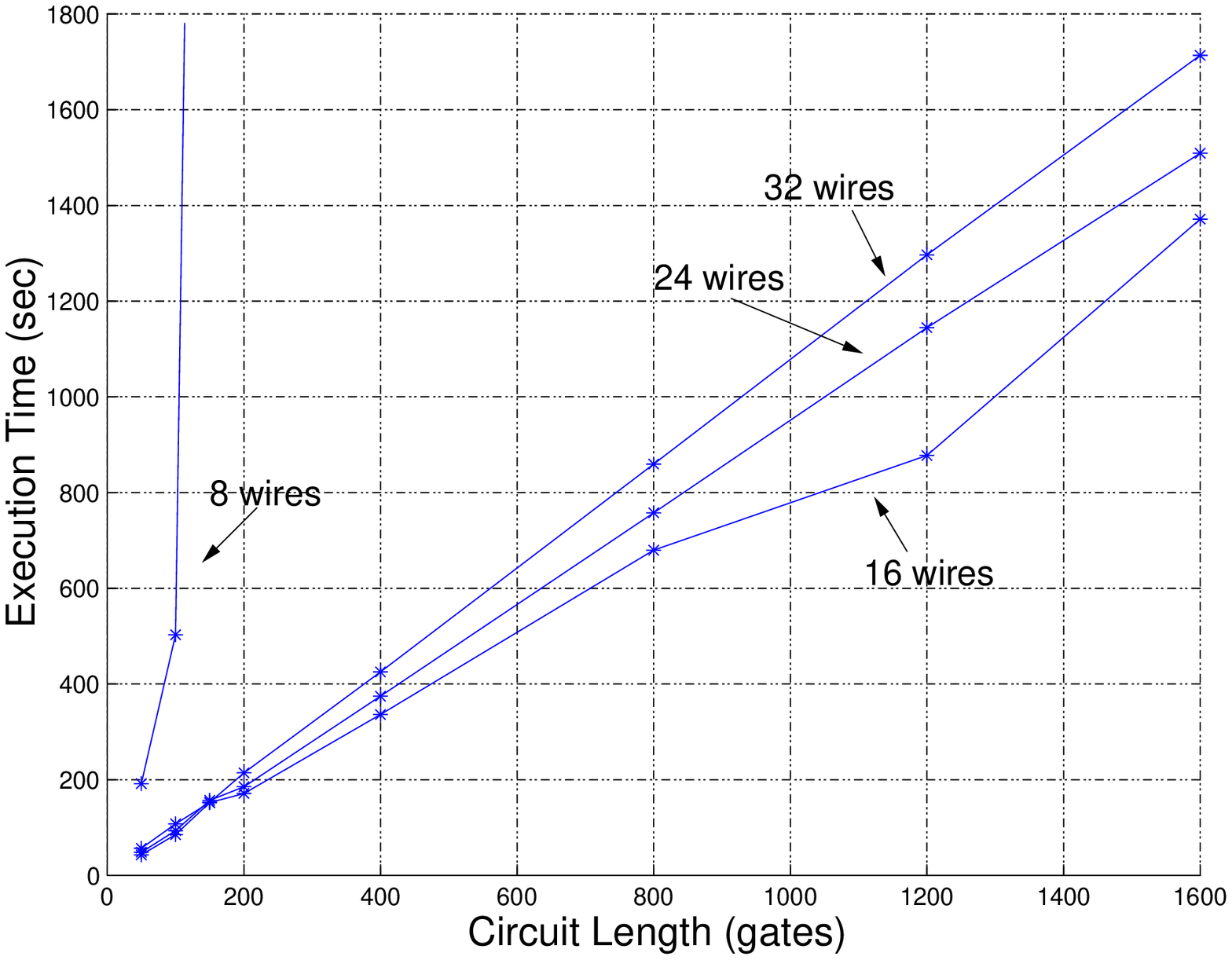}}\\
\mbox{(a)} & \mbox{(b)}
\end{array}$ 
%    {\includegraphics{test_size.eps}}
\caption{Simulations results for circuit decomposition algorithm
limiting sub-circuit size to 8 wires. (a) Average test set size
(after compaction) versus circuit length.
Staircase graph represents the upper bound given in
Proposition~\ref{prop:constr}c. (b) Execution time versus circuit length.}\label{fig:test_size}  
\end{center}
\end{figure}
We conducted a set of simulations to evaluate the performance of our algorithm.
We generated random $NCT$-circuits of various lengths over 8, 16, 24 and 32 
wires. The circuits were generated by selecting at random from the 
set of all allowable NOT, C-NOT, and Toffoli gates. Each circuit was 
decomposed into sub-circuits acting on at most 8 wires, and our algorithm was
used to find a complete test set. 
Figure~\ref{fig:test_size}a shows the average number of test vectors needed
as a function of the circuit length. At least 150 circuits were generated for
each data point.

The average execution time for the algorithm
%, shown in 
%Figure~\ref{fig:ex_time}, 
seems to increase linearly with circuit length
and does not vary very much with the number of input/output wires,
with the exception of the 8-wire case for which execution time appears
to increase exponentially with circuit
length~(Figure~\ref{fig:test_size}b). This latter case is most 
likely because the number of constraints increases linearly 
with the number of gates, yielding  increasingly difficult ILPs. On the other hand, 
for the circuits on more than 8 wires, an increase in the length of circuit does not 
generally lead to significantly harder individual ILPs, rather only a (linearly) larger number 
of them to solve. 

Test compaction, as expected, is most effective for longer circuits, 
eliminating an average of approximately one redundant test vector for 
circuits containing $800$ or more gates.
%\begin{figure}[t!]
%  \begin{center}
%    \resizebox{3.5in}{!}
%    {\includegraphics{ex_time.eps}}
%\caption{Average execution time vs. circuit length for circuit decomposition algorithm limiting sub-circuit size to 8 wires.}\label{fig:ex_time} 
%  \end{center}
%\end{figure}

\subsection{Comparison to Conventional ATPG}
\begin{figure}[t!]
\begin{center}
$\begin{array}{cc}
\resizebox{3.1in}{!}
{\includegraphics{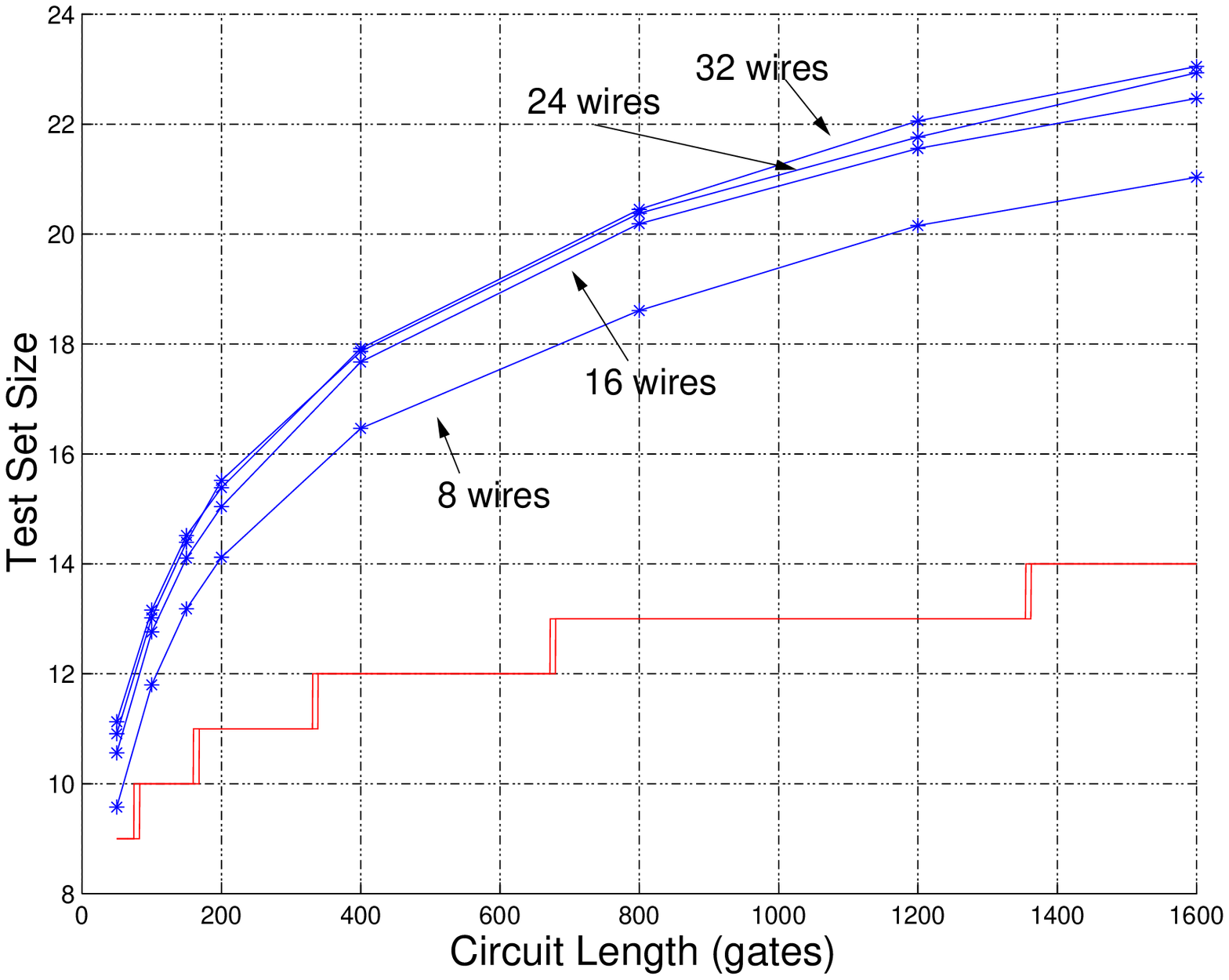}} &
\resizebox{3.1in}{!} 
{\includegraphics{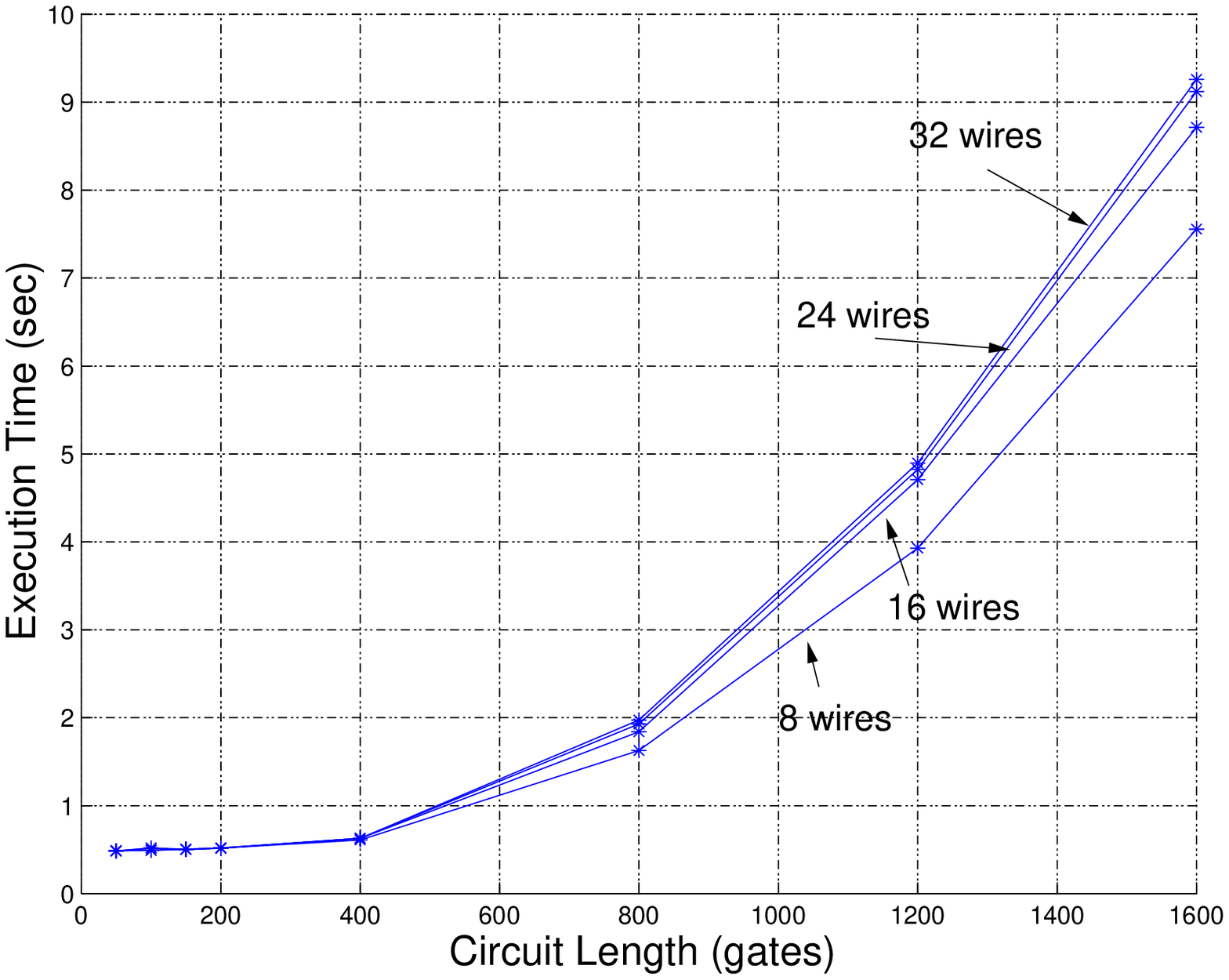}} \\ 
%    {\includegraphics{test_size.eps}}
\mbox{(a)} & \mbox{(b)}
\end{array}$ 
\caption{Simulation results for Atalanta~\cite{atalanta}. (a) Average
test set size versus circuit length. Staircase graph represents the
upper bound given in Proposition~\ref{prop:constr}c. (b)
Execution time versus circuit length.}\label{fig:ATPG}  
\end{center}
\end{figure}
A number of ATPG software packages are available for generating
test sets for conventional combinational circuits, and some of these
can be readily modified for the reversible case.  Here we used the
ATPG tool Atalanta~\cite{atalanta}, because of 
its ease of use and the availability of its source code. Since
the Toffoli gate used in our reversible circuits is not a standard
combinational logic gate, we had to make some minor modification to the code
to handle this gate.  Basically, we replaced each Toffoli gate by an
equivalent combinational circuit using conventional irreversible
gates, and modified the code to ignore faults in the internal
nodes of these sub-circuits.

Figure~\ref{fig:ATPG} shows the average size of test sets generated by
Atalanta as a function of the number of gates in the reversible
circuit. Results are shown for 8, 16, 24 and 32 input/output wires. As
the figure illustrates, the test sets given by Atalanta are, on average,
almost twice as large as those given by our circuit decomposition algorithm,
and their average size is greater than the upper bound of
Proposition~\ref{prop:constr}c.  However, Atalanta is significantly
faster than our algorithm requiring an average
of less than 10 seconds for circuits with 32 wires and 1600 gates;
this compares to approximately 30 minutes for the circuit
decomposition algorithm.  However, the execution time for Atalanta
appears to increase at a much faster rate with respect to the circuit
length than that of the circuit decomposition algorithm.

\section{Cell Fault Model}\label{sec:cell_fault}
While the use of the stuck-at fault model has been very
effective in conventional circuit testing, other fault models 
may be more appropriate for reversible circuits, especially in the
quantum domain. For example, the cell fault model~\cite{kautz:tff:67}, 
where the function of the faulty $k\times k$ gate 
changes arbitrarily from the desired function, may be more realistic.
In this section we extend some of our results to this model.

The following proposition provides a basic necessary and sufficient 
condition for a test set to be complete for the cell fault model. While this 
condition is also necessary for irreversible circuits, in that latter 
case it is not sufficient.
\begin{proposition}\label{prop:suff_comp2}
Under the cell fault model a test set is complete if and 
only if the inputs of every $k\times k$ gate in the circuit can be set
to all $2^k$ possible values by the test set.
\end{proposition}
\textbf{Proof}
If a test set does not set the input wires of a gate to a 
particular value say $\vect{a}$, then it would not be 
able to detect a failure in this gate that only affects the output of $\vect{a}$.
On the other hand, if the input wires of every gate in the 
circuit can be set to all possible values by the test set, then any 
single-gate failure will affect at least one test vector, changing 
the value at the output of the gate. By the observability property
of reversible circuits, this will be reflected in a change at the output. $\Box$

\noindent
As an example, consider a circuit with a C-NOT gate.  In order to detect any 
fault in the C-NOT gate the test set should be able to set the inputs of the 
gate to $\{00,\ 01,\ 10, \ \mbox{and} \ 11\}$.  If the gate is faulty, it will 
operate incorrectly on at least one of these input values which will then be 
reflected in an incorrect circuit output.

Let $g_1,\ldots,g_l$ be the gates in a reversible circuit, 
and $k_1,\ldots,k_l$ the respective gate sizes.  
If we consider every possible value at the input of each gate as 
representing a distinct fault, the total number of faults that need to
be covered is 
$\sum_{i=1}^l 2^{k_i}$. Under this definition, we have the following lemma.
\begin{lemma}\label{lemma:count2}
Each input vector covers exactly $l$ faults, and a fault associated with 
a $k\times k$ gate is covered by exactly $2^{n-k}$ input vectors.
\end{lemma}
\textbf{Proof}
Each input vector sets the bits at the inputs of each gate to some value. 
Therefore, since there are $l$ gates, the vector can detect $l$ faults.
For a given fault associated with a $k\times k$ gate there are 
$2^{n-k}$ possible values for the $n$ bits at that level that can detect 
it. Since the circuit is reversible, each of these can be traced back to 
a distinct input vector.
$\Box$

The following proposition, which is analogous to Proposition~\ref{prop:constr},
gives upper bounds on the size of the minimal test set under the
cell fault model.
\begin{proposition}\label{prop:cf_bounds}
A complete test set under the cell fault model for an $n$-wire reversible 
circuit with a total of $l$ gates with sizes $k_1 \ge k_2 \ge \ldots \ge k_l$ is given by
%\footnote{Any gate with fewer than
%$k$ inputs/outputs can be converted into a $k \times k$ gate by passing unused 
%wires through the gate. Any test set that fully tests the new gate must also 
%fully test the original.}:
\begin{itemize}
\item[a.] any $2^{n}-2^{n-k_1}+1$ distinct test vectors
\item[b.] a set of $\left(\sum_{i=1}^{l}2^{k_i}\right)-l+1$ test vectors
\item[c.] some set of at most $\sum_{i=1}^{l}\left\lceil \frac{2^{k_i}}{i}\right\rceil$ test vectors
\end{itemize}
\end{proposition}
\textbf{Proof}\\
(a) For any $k\times k$ gate in the circuit there are $2^{n-k}$ distinct 
inputs that yield a particular value at its input. Therefore, if the test set
has $2^{n}-2^{n-k_1}+1$ vectors (implying that fewer than $2^{n-k}$ are not 
included) then it must include at least one such input. Since this is true for all gates in the circuit, by 
Proposition~\ref{prop:suff_comp2}, the test set is complete.\\
(b) Any input vector will cover $l$ faults leaving 
$\sum_{i=1}^{l}2^{k_i}-l$. By the controllability property we can cover 
these with one test vector each. Therefore, all of the faults can be covered
with no more than $\sum_{i=1}^{l}2^{k_i}-l+1$ test vectors. \\
(c) We first prove that given an incomplete set of $m$ test vectors covering 
faults in the set $F_C$, there must be a test vector that covers at least 
\begin{equation} \label{eqn:propc}
  l-\left\lfloor \sum_{f\in F_C} {2^{-k(f)}} \right\rfloor
\end{equation}
of the remaining faults, where $k(f)$ is the size of the gate associated
with fault $f$.

Suppose this is false. By Lemma~\ref{lemma:count2} every test vector covers exactly $l$ faults and a fault $f$
is covered by exactly $2^{n-k(f)}$ input values. Therefore the number of times faults in $F_C$ can be covered is $\sum_{f\in F_C} 2^{n-k(f)}$ and the current 
test set accounts for $m l$ of these. Furthermore, each of the remaining input vectors 
must cover more than $\sum_{f\in F_C} 2^{-k(f)}$ of the already covered faults, otherwise our 
assertion would be true. Combining these we have the following inequalities.
\begin{eqnarray}
%\left(2^n-m\right)\left(\left\lceil \sum_{f\in F_C} 2^{-k(f)}\right\rceil\right) & \le & \sum_{f\in F_C}2^{n-k(f)}-m\cdot l \\
\left(2^n-m\right)\left(\sum_{f\in F_C} 2^{-k(f)}\right) & < & \sum_{f\in F_C}2^{n-k(f)}-m\cdot l \\
l &<& \sum_{f\in F_C} 2^{-k(f)}
\end{eqnarray}
The second inequality is false since the right side can be no larger than $l$.
Therefore, we have a contradiction, and our proposition must be true.

Iteratively removing $l-\left\lfloor \sum_{f\in F_C} {2^{-k(f)}} \right\rfloor$ faults 
from the set of uncovered faults eventually leaves the set empty.  The number of iterations
needed to do this is an upper bound on the number of test vectors needed for completeness. The 
floor function in the equation makes it difficult to obtain a closed form for the bound, but we can
weaken the above result to do this.  To cover the first $2^{k_l}$ faults we need at most 
$\lceil 2^{k_l}/l\rceil$ test vectors, since each test vector we add covers $l$ faults. To cover 
the next $2^{k_{l-1}}$ faults we need at most $\lceil 2^{k_{l-1}}/(l-1)\rceil$ test vectors, and 
so on. Thus, we can cover all single cell faults using no more than 
$\sum_{i=1}^{l}\left\lceil 2^{k_i}/i\right\rceil$ test vectors. 
$\Box$

%This proposition does not yield very tight upper bounds on the test set size.
For a reversible circuit on 64 wires with a million
$3\times 3$ gates, parts a-c of Proposition~\ref{prop:cf_bounds} give 
upper bounds of approximately
$10^{19}$, $7\cdot 10^6$ and $10^6$ test vectors,
respectively. However, since part c uses the property illustrated in
Equation~(\ref{eqn:propc}) very conservatively in order 
to obtain a closed form, a much tighter bound can be obtained by applying this
property directly. In fact, by iteratively applying this property, one
can show that no more than $108$ test vectors are needed for complete
testing. In general we can approximate this tighter bound if we assume
that the number of faults covered after any given iteration when taken mod
$2^k$ is uniformly distributed on the integers 0 through $2^k-1$
inclusively. The expected value of~(\ref{eqn:propc}) is then
\begin{equation}
 \mbox{E}\left[l-\left\lfloor \sum_{f\in F_C} {2^{-k(f)}}
    \right\rfloor\right] = \mbox{E}\left[l -
    \left\lfloor\frac{\left|F_C\right|}{2^{k}}\right\rfloor\right] = 
    l-\frac{\left|F_C\right|}{2^{k}}+\frac{2^k-1}{2^{k+1}}
\end{equation}
This then gives the following recursion on $C(n)$, the minimum number of faults that
can be covered by $n$ test vectors:
\begin{eqnarray}
 C(n) &= & C(n-1) + \left(l - \frac{
 C(n-1)}{2^{k}}+\frac{2^k-1}{2^{k+1}}\right)\\
      &= & \frac{2^k-1}{2^k}\cdot C(n-1) + l + \frac{2^k-1}{2^{k+1}}
\end{eqnarray}
Letting $a = (2^k-1)/2^k$ and $b = l + (2^k-1)/2^{k+1}$,
\begin{eqnarray}
 C(n) &= & a^2 \cdot C(n-2) + a \cdot b + b\\
      &= &  a^n \cdot C(0) +  \left(a^{n-1} + a^{n-2} + \cdots + 1
      \right) \cdot b \\
      &= & \frac{1-a^n}{1-a} \cdot b
      \ \ \ = \ \ \  2^k\cdot
 \left(1-\left(\frac{2^k-1}{2^k}\right)^n\right) \cdot \left(l +
 \frac{2^k-1}{2^{k+1}}\right)
\end{eqnarray}
In order to cover all faults, $C(n)$ must be greater than or equal to
the total number of faults: 
\begin{equation}
  C(n) = 2^k\cdot \left(1-\left(\frac{2^k-1}{2^k}\right)^n\right)
 \cdot \left(l + \frac{2^k-1}{2^{k+1}}\right) \ge l \cdot 2^k 
\end{equation}
On rearranging we have 
\begin{equation}
%   \left(1-\left(\frac{2^k-1}{2^k}\right)^{n-1}\right)
% \cdot \left(l + \frac{2^k-1}{2^{k+1}}\right) & \ge & l \\
% \frac{2^k-1}{2^{k+1}} & \ge & \left(l + \frac{2^k-1}{2^{k+1}}\right)
% \cdot \left(\frac{2^k-1}{2^k}\right)^{n-1}\\ 
% \left(\frac{2^k}{2^k-1}\right)^{n-1} & \ge & \left(
% \frac{2^{k+1}}{2^k-1}\cdot l +1 \right)\\
 n \ge \frac{\log_2\left(\frac{2^{k+1}}{2^k-1}\cdot l + 1
 \right)}{k - \log_2\left(2^k-1\right)}
\end{equation}
This result strongly suggests the number of test vectors needed for
completeness under the cell fault model grows logarithmically in the
number of gates, just as in the stuck-at fault case.   

\begin{figure}[t!]
  \begin{center}
  $\begin{array}{cc}
    \resizebox{3.1in}{!}
    {\includegraphics{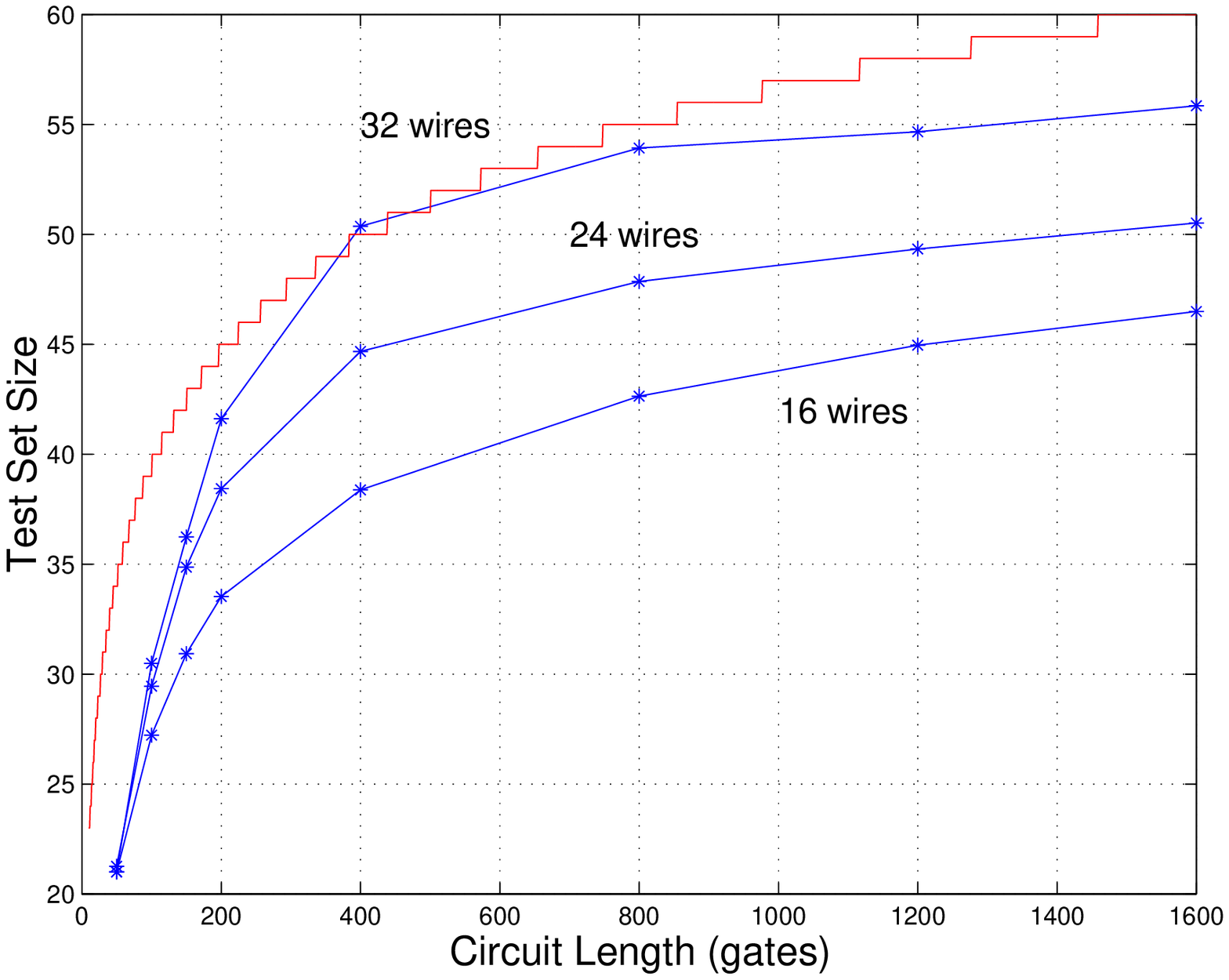}} &
    \resizebox{3.1in}{!}
    {\includegraphics{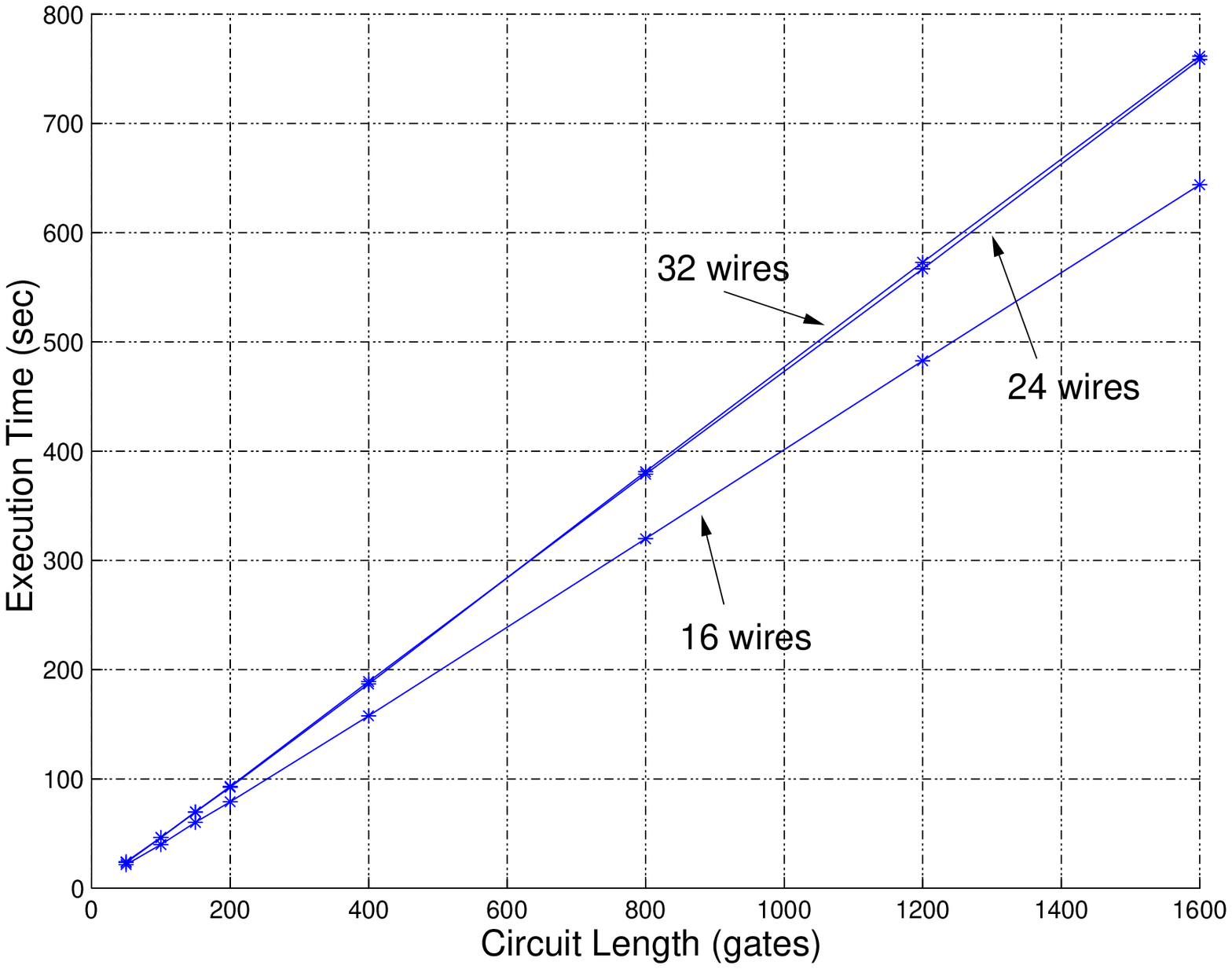}} \\
   \mbox{(a)} & \mbox{(b)}
\end{array}$
%    {\includegraphics{test_size.eps}}
\caption{Simulation results for cell fault model. (a) Average test set
    size (after compaction) vs. circuit length for circuit  
    decomposition algorithm using the cell fault model. The
    sub-circuit sizes are limited to 8 wires in the
    decomposition. The staircase graph represents the upper bound
    given by iterating the property in~(\ref{eqn:propc}). (b)
    Execution time versus circuit length.}\label{fig:cf_test_size}  
  \end{center}
\end{figure}

To obtain an ILP formulation for the cell fault model only the  
constraints given in Section~\ref{sec:ilp} need to be modified. For each
$k\times k$ gate at each level we generate $2^k$ constraints, one for each of
the possible inputs to the gate. 
The circuit decomposition method
from Section~\ref{subsec:circ_decomp} as well as the test set compaction 
method in Section~\ref{subsec:test_comp} can be applied as in the 
stuck-at fault case.  Figure~\ref{fig:cf_test_size} shows simulation
results for the circuit decomposition algorithm under the cell fault
model. The staircase graph represents the upper bound obtained by
iterating~(\ref{eqn:propc}).  The average size of the generated test
set is generally below this bound.  Test set compaction results in
the elimination of 4-5 test vectors on average, with more vectors
eliminated for longer circuits and fewer for shorter ones.     

Empirically, the algorithm requires half the execution time
of the stuck-at fault case, with the exception of the 8-wire case for
which we were unable to obtain results due to prohibitive execution
times.

\section{Conclusions}\label{sec:concl}
We have considered the test set generation problem for reversible
circuits, and shown that the property of reversibility fundamentally 
simplifies the problem.  For example, a test set that detects all single 
stuck-at faults in a reversible circuit also detects all multiple faults. 
We have derived test set completeness conditions under both the stuck-at 
and cell fault models. We have then used these to find general test set constructions 
that implicitly yield upper bounds on the number of test vectors needed for completeness.  
One bound shows that the test set size increases at most logarithmically with the 
length of circuit, strengthening  our assertion that reversible circuits are easier 
to test than conventional ones.  We have also given a practical algorithm for finding 
complete test sets. Our algorithm generates test sets that are approximately half 
the size of those produced by conventional ATPG. 

In addition to the fault detection problem we have investigated here, we
also plan to study fault diagnosis,  
that is, using test sets to localize faults. As with the detection problem, fault diagnosis 
may be easier for reversible circuits than for irreversible ones.
Finally, though we have focused on testing for classical reversible circuits, 
we also hope to extend our work to the quantum case. The latter is likely to be very
different from the former: while fault-free classical 
circuits are deterministic, fault-free quantum ones are inherently probabilistic.
Thus, the goal for the quantum case may be to determine as efficiently as possible, and
with a given degree of confidence, whether the circuit contains a fault or not.  A step
towards this goal may be to find a small set of test vectors that sufficiently exercises
the internal gates in the circuit.  Our results for the cell fault model studied in 
Section~\ref{sec:cell_fault} may be particularly useful for this.

%\bibliography{FT_ref}

\begin{thebibliography}{10}

\bibitem{agrawal:ait:81}
V.~D. Agrawal.
\newblock An Information Theoretic Approach to Digital Fault Testing.
\newblock {\em IEEE Transactions on Computers}, vol. 30, pp. 582--587, August 1981.

\bibitem{bennett:lro:73}
C.~H. Bennett.
\newblock Logical Reversibility of Computation.
\newblock {\em IBM Journal of Research and Development}, vol. 17, pp. 525--532,
November 1973.

\bibitem{bertrand:slr:74}
J.~C. Bertrand, N.~Giambiasi, and J.~J. Mercier.
\newblock Sur la Recherche de l'Inverse d'un Automate.
\newblock {\em RAIRO}, pp. 64--87, April 1974.

\bibitem{bertrand:slr2:74}
J.~C. Bertrand, J.~J. Mercier, and N.~Giambiasi.
\newblock Sur la Recherche de l'Inverse d'un Circuit Combinatoire.
\newblock {\em RAIRO}, pp. 21--44, July 1974.

\bibitem{buhrman:tas:01}
H.~Buhrman, J. Tromp, and P.~Vit\'{a}nyi.
\newblock Time and Space Bounds for Reversible Simulation.
\newblock {\em Journal of Physics A: Mathematical and General}, vol. 34,
pp. 6821--6830, September 2001. 
 
\bibitem{bushnell:eoe:00}
M.~L. Bushnell and V.~D. Agrawal.
\newblock {\em Essentials of Electronic Testing for Digital Memory \&
  Mixed-Signal VLSI Circuits}.
\newblock Kluwer Academic Publishers, Boston, 2000.

\bibitem{chang:tsc:95}
J.-S. Chang and C.-S. Lin.
\newblock Test Set Compaction for Combinational Circuits.
\newblock {\em IEEE Transactions on CAD}, vol. 14, pp. 1370--1378, November
  1995.

\bibitem{debany:afg:91}
W.~H. Debany Jr., C~.R.~P. Hartmann and T.~J. Snethen.
\newblock Algorithm for Generating Optimal Tests for Exclusive-OR Networks.
\newblock {\em IEE Proceedings E (Computers and Digital
  Techniques)}, vol. 138,  pp. 93--96, March 1991.

\bibitem{devos:arc:02}
B.~Desoete and A.~De~Vos.
\newblock A Reversible Carry-Look-Ahead Adder Using Control Gates.
\newblock {\em Integration, The VLSI Journal}, vol. 33, pp. 89--104, 2002.

\bibitem{flores:oas:99}
P.~F. Flores, H.~C. Neto, and J.~P. Marques-Silva.
\newblock On Applying Set Covering Models to Test Set Compaction.
\newblock {\em Proceedings of GLS-VLSI}, pp. 8--11, March 1999.

\bibitem{fredkin:cl:82}
E.~Fredkin and T.~Toffoli.
\newblock Conservative Logic.
\newblock {\em Intl. Journal of Theoretical Physics}, vol. 21,
pp. 219--253, 1982. 

\bibitem{garey:79:cai}
M.~R. Garey and D.~S. Johnson.
\newblock {\em Computers and Intractability: A Guide to the Theory of
  NP-Completeness}.
\newblock W. H. Freeman and Company, New York, 1979.

\bibitem{goel:tgd:79}
P.~Goel and B.~C. Rosales.
\newblock Test Generation \& Dynamic Compaction of Test.
\newblock {\em Digest of Papers Test Conference}, pp. 189--192,
October 1979. 

\bibitem{hayes:oro:71}
J.~P. Hayes.
\newblock On Realizations of Boolean Functions Requiring a Minimal or
Near Minimal Number of Tests.
\newblock {\em IEEE Transactions on Computers}, vol. 20,
pp. 1506--1513, December 1971. 

\bibitem{hochbaum:aot:96}
D.~S. Hochbaum.
\newblock An Optimal Test Compression Procedure for Combinational Circuits.
\newblock {\em IEEE Transactions on CAD}, vol. 15, pp. 1294--1299, October 1996.

\bibitem{cplex}
ILOG CPLEX.
\newblock http://www.ilog.com/products/cplex.

\bibitem{kautz:tff:67}
W.~H. Kautz.
\newblock Testing for Faults in Cellular Logic Arrays.
\newblock {\em Annual Symposium on Switching and Automata Theory}, pp.
  161--174, 1967.

\bibitem{landauer:iah:61}
R.~Landauer.
\newblock Irreversibility and Heat Generation in the Computing Process.
\newblock {\em IBM Journal of Research and Development}, vol. 3, pp. 183--191,
July 1961. 

\bibitem{li:rso:98}
M.~Li, J. Tromp, and P. Vit\'{a}nyi.
\newblock Reversible Simulation of Irreversible Computation.
\newblock {\em Physica D}, pp. 168--176, September 1998.
 
\bibitem{atalanta}
H.~K.~Lee and D.~S.~Ha. 
\newblock On the Generation of Test Patterns for Combinational
Circuits.
\newblock {\em Technical Report No. 12\_93}, 
\newblock Dept. of Electrical Engineering, 
\newblock Virginia Polytechnic Institute and State University.

\bibitem{nielsen:qca:00}
M.~A. Nielsen and I.~L. Chuang.
\newblock {\em Quantum Computation and Quantum Information}.
\newblock Cambridge University Press, 2000.

%\bibitem{shende:rlc:02}
%V.~V. Shende, A.~K. Prasad, I.~L. Markov, and J.~P. Hayes.
%\newblock Reversible Logic Circuit Synthesis.
%\newblock {\em Proceedings of IEEE/ACM Intl. Conf. on CAD},
%pp. 353--360, November 2002.

\bibitem{shende:sor:03}
V.~V. Shende, A.~K. Prasad, I.~L. Markov, and J.~P. Hayes.
\newblock Synthesis of Reversible Logic Circuits.
\newblock {\em IEEE Transactions on CAD}, vol. 22, 
pp. 710--722, June 2003.

\bibitem{toffoli:rc:80}
T. Toffoli.
\newblock Reversible Computing.
\newblock {\em Automata, Languages and Programming, 7th Colloquium}, 
J.~W. de Bakker and J.~van Leeuwen (eds.),
\newblock Lecture Notes in Computer Science No. 85, Springer-Verlag, 
pp. 632-644, 1980.

\bibitem{zhirnov:ltb:03}
V.~V.~Zhirnov, R.~K. Calvin III, J.~A. Hutchby, and G.~I. Bourianoff.
\newblock Limits to Binary Logic Switch Scaling---A Gedanken
  Model.
\newblock {\em Proceedings of the IEEE}, vol. 91, pp. 1934--1939, 
November 2003.  
% \bibitem{stephan:ctg:96}
% P.~Stephan, R.~K. Brayton, and A.~L. Sangiovanni-Vincentelli.
% \newblock Combinational Test Generation Using Satisfiability.
% \newblock {\em IEEE Transactions on CAD}, vol. 15, pp. 1167--1176,
% September 1996. 

\end{thebibliography}
\end{document}